\documentstyle[12pt]{article}

\title{
\begin{flushright}
{\normalsize Yaroslavl State University\\
             Preprint YARU-HE-97/08\\
             hep-ph/9712289} \\[5mm]
\end{flushright}
Photon decay $\gamma \to \nu \bar \nu$ in
       an external magnetic field}

\author{A.V.~Kuznetsov$^{1)}$, N.V.~Mikheev$^{1)}$ and 
L.A.~Vassilevskaya$^{1,2)}$\\ [7mm] 
       {\small\it 1) Department of Theoretical Physics, 
               Yaroslavl State University,} \\
       {\small\it    Sovietskaya 14, Yaroslavl 150000, 
               Russia} \\
       {\small\it  (E-mail addresses: avkuzn@uniyar.ac.ru, 
               mikheev@yars.free.net)} \\
       {\small\it 2) Moscow Lomonosov University, V-952, Moscow 117234, 
               Russia} \\
        {\small\it   (E-mail address: vasilevs@vitep5.itep.ru)}}

\date{7 December 1997, revised 20 February 1998}

\begin{document}

\maketitle

\begin{abstract}
The process of the photon decay into the neutrino - antineutrino pair 
in a magnetic field is investigated. The amplitude and the probability are 
analysed in the limits of relatively small and strong fields. 
The probability is suppressed by a factor $(G_F \;m^2_e)^2$ 
as compared with the probability of the pure electromagnetic 
process $\gamma \to e^- e^+$. However, the process with neutrinos could 
play a role of an additional channel of stellar energy-loss.
\end{abstract}

\vspace{10mm}

\centerline{\it Submitted to Physics Letters B}

Nowadays, it is generally recognized that astrophysical objects and 
pro\-ces\-ses 
inside them give us unique possibilities~\cite{Raf1} for investigations of 
particle properties under extreme conditions of a high density and/or 
temperature of matter, and also of a strong magnetic field. A concept of the 
astrophysically strong magnetic field has been changed in the recent years 
and now the field is considered as the strong one 
if it is much greater than the known Schwinger value, 
$B \gg B_e, B_e = m^2_e/e \simeq 4.41 \cdot 10^{13} G$. 
Possible mechanisms are now discussed of a generation of such strong fields 
($B \sim 10^{15} - 10^{17} G$) in astrophysical cataclysms like a supernova 
explosion or a coalescence of neutron stars~\cite{tor,pol}, and in the early 
Universe~\cite{cosm}. 

The strong magnetic field, like the medium, makes an active influence 
on particle properties. First, it could induce new interactions, for example, 
an interaction of the uncharged massless neutrinos with photons. Second, 
the field changes essentially the kinematics of particles, causing new 
channels to be opened, which are forbidden in vacuum by the momentum 
conservation:  $\gamma \to e^- e^+$~\cite{Klep}, $\nu \to \nu \gamma$ 
~\cite{GN,Skob,IR,GMV}, $\nu \to \nu e^- e^+$~\cite{Bor,KM}.

In this paper, the process of the photon decay into the neutrino - 
antineutrino pair $\gamma \to \nu \bar \nu$ is investigated in a presence 
of an external magnetic field. The field plays a role of the active medium 
to change the photon dispersion in such a way that the states with the 
timelike photon 4-momentum $q$ ($q^2 > 0$) become possible. Let us note that 
plasma influences the photon properties similarly. The process 
$\gamma \to \nu \bar \nu$ in plasma is well studied and has an importance 
in astrophysics as one of the main sources of stellar cooling 
(see a detailed review in the book~\cite{Raf1}).

It seems reasonable to say that both components of the active medium, a 
magnetic field and plasma, are presented in the most of astrophysical objects. 
A situation is also possible when the magnetic component dominates. 
For example, in a supernova explosion or in a coalescence of neutron 
stars a region could exist outside the neutrinosphere, of order of several 
hundred kilometres in size, where plasma is rather rarefied, and the magnetic 
field of the toroidal type could reach a value of 
$10^{14} - 10^{16} G$~\cite{tor}.

The purpose of our work is to calculate the probability of the photon 
decay $\gamma \to \nu \bar \nu$ in a strong magnetic field 
in the frame of the standard model, 
and to evaluate a contribution of this process into the neutrino luminosity 
in the above-mentioned astrophysical cataclysm. 

It is known that two eigenmodes of the real photon exist in a magnetic 
field~\cite{Ad} with polarization vectors

\begin{equation}
\varepsilon _{\mu}^{(\parallel)} = 
\frac{ (q \varphi)_{\mu} }{ \sqrt{ 
 q^2_\perp  } } ; \; \; \; \; \;
\varepsilon _{\mu}^{(\perp)} = \frac{ (q \tilde
\varphi)_{\mu} }{ \sqrt{ q^2_\parallel } },
\label{eq:eps}
\end{equation}

\noindent where $ \varphi_{\alpha \rho} = F_{\alpha \rho} / B$
is the dimensionless tensor of the external magnetic field,  
${\tilde \varphi}_{\alpha \rho} = \frac{1}{2} \varepsilon_{\alpha \rho
\mu \nu} \varphi_{\mu \nu} \; $ is the dual tensor, 
$q^2_{\parallel}  =  ( q \tilde \varphi \tilde \varphi q ) =
q_\alpha \tilde \varphi_{\alpha\rho} \tilde \varphi_{\rho\mu} q_\mu$,
$q^2_{\perp}  =  ( q \varphi \varphi q )$. 
For the relatively small ``photon mass'', $|q^2| \ll \omega^2$, one can see 
that 
$q^2_{\parallel} \simeq q^2_{\perp} \simeq \omega^2 \sin^2 \theta$, where
$\omega$ is the photon energy,
$\theta$ is the angle between the photon momentum
$\vec q$ and the magnetic field direction. 
The vectors~(\ref{eq:eps}) are the eigenvectors of the photon polarization 
tensor in a magnetic field. The photon decay $\gamma \to \nu \bar \nu$, 
Fig. 1, is the crossed channel of the gamma radiation by a neutrino 
$\nu \to  \nu \gamma $. 
Due to the photon dispersion in a magnetic field~\cite{Shab},
the necessary condition of the decay, $q^2 > 0$, is
realized for the perpendicular ($\perp$) polarization in the region 
$q^2_\parallel > 4 m_e^2$ and for the parallel
($\parallel$) polarization in the region 
$q^2_\parallel > ( m_e + \sqrt{m^2_e + 2 e B})^2$.

However, due to the collinearity of the kinematics,
$j_{\alpha} \sim q_{\alpha} \sim p_{\alpha} \sim p'_{\alpha}$, 
the amplitude of the $\parallel$ mode decay is suppressed, and for the 
$\perp$ mode it takes the form

\begin{equation}
M_{\perp} \simeq \frac{ 2 \; e \; G_F \; g_A}{\sqrt{2} \pi^2} \;
\sqrt{x (1 - x)} \; [e^2 (q F F q)]^{1/2} \; J,
\label{eq:M1}
\end{equation}

\noindent where $e > 0$ is the elementary charge, $g_A$ is the axial-vector 
constant in the neutral electron current in the effective $\nu \nu e e$ 
Lagrangian, $x = E/\omega$, 

\noindent 
$1 - x = E '/\omega$ are the relative energies 
of the neutrino and the anti-neutrino correspondingly.
The neutrino current $j_{\alpha}$ is substituted into the 
amplitude~(\ref{eq:M1}) in the form

\begin{equation}
j_{\alpha} = \bar \nu (p) \gamma_{\alpha} (1 - \gamma_5) \nu(- p ')
\simeq 4 \sqrt{x (1 - x)}\; q_\alpha, 
\label{eq:j} 
\end{equation}

\noindent where the collinearity of the kinematics is taken into account. 
The amplitude~(\ref{eq:M1}) describes the process of the photon decay 
into the electron-neutrino pair ($g_A = + 1/2$) and into the 
muon-neutrino and tau-neuitrino pairs ($g_A = - 1/2$) as well. 

The dimensionless formfactor $J$ in the general case has a form of a double 
integral

\begin{equation}
J = 1 - i \;m^2_e \; \int\limits_0^1 d u \int\limits_0^\infty d t \;
exp \left \{ -i \left [
t \left ( m^2_e -  q^2_{\parallel}  \frac{ 1 - u^2 }{4} \right ) +  
\frac{q^2_{\perp}}{2 \beta} \; \frac{ \cos \beta u t - \cos \beta t }
{ \sin \beta t }
\right ] \right \},
\label{eq:J1} 
\end{equation}

\noindent where $\beta = e B$, and it can be calculated rather easily in two 
limiting cases.

 i) If the field strength $B$ appears to be the largest physical parameter,
 $e B ~\gg ~q^2_{\parallel}$, one obtains

\begin{equation}
J \simeq  \frac{1 - v^2 }{2 v}
\left ( 
ln \; \frac{ 1 + v }{1 - v } - i \pi
\right ) + 1,
\label{eq:J2}
\end{equation}

\noindent where $v = \sqrt{1 - 4 m^2_e/q^2_{\parallel}}$.

 ii) In the alternative case, when $e B ~\ll ~q^2_{\parallel}$, 

\begin{equation}
J \simeq 1.
\label{eq:J3}
\end{equation}

The total probability of the photon decay into all neutrino species 
is defined by the expression

\begin{equation}
W_{\perp} = {1 \over 16 \pi \omega} \,
\int\limits^{1}_{0} dx \; | M_{\perp} |^2 = 
\frac{\alpha \; G_F^2}{16 \pi^4 \omega} \; 
e^2 (q F F q) \; | J |^2.
\label{eq:W1}
\end{equation}

\noindent Here we assume that all neutrino masses are much smaller than the 
field-induced ``photon mass''.
Let us note that the photon decay probability~(\ref{eq:W1}) 
with~(\ref{eq:J2}) contains at first sight the pole-type singularity at 
 $q^2_\parallel  \to 4 m^2_e $ ($v \to 0$). However, the solution 
of the equation of the photon dispersion in this limit shows that

\begin{equation}
| q^2_\parallel - 4 m^2_e |_{min} = \omega \; 
\Gamma_{\gamma \to e^- e^+}.
\label{eq:disp}
\end{equation}

\noindent It is known that the similar seeming singularity, but of the 
square-root-type, takes place in the process of the photon decay into 
the electron-positron pair $\gamma \to e^- e^+$~\cite{Klep}. As was shown 
in Ref.~\cite{Shab}, taking account of the photon dispersion in the process 
$\gamma \to e^- e^+$ leads to a finite value for the decay width, which is 
maximal at the point $q^2_\parallel = 4 m^2_e$

\begin{equation}
\left ( \Gamma_{\gamma \to e^- e^+} \right)_{max} = 
\frac{\sqrt{3}}{2} \; \left (
\frac{2 \alpha e B}{m^2_e} \right)^{2/3} \; \frac{m^2_e}{\omega}.
\label{eq:max}
\end{equation}

\noindent 
The probability of the decay $\gamma \to \nu \bar \nu$ is also finite, 
in view of Eqs.~(\ref{eq:disp}) and~(\ref{eq:max}), and amounts up to the 
maximal value:

\begin{equation}
(W_{\perp})_{max} = \frac{(G_F \;m^2_e)^2}
 {4 \sqrt{3} \pi^2 } \,
\left (\frac{2 \alpha e B}{m^2_e} \right)^{1/3} \; 
\frac{e B}{\omega}.
\label{eq:W2}
\end{equation}

\noindent In the limiting case of a rather weak field 
$e B ~\ll ~q^2_{\parallel}$ the probability~(\ref{eq:W1}) with~(\ref{eq:J3}) 
coincides with the result of Ref.~\cite{GN}, to the factor 3/4. 
This discrepancy is due to the fact that the contribution of the $Z$ boson 
was not taken into account in Ref.~\cite{GN}.

It is evident that the probability~(\ref{eq:W2}) of the electroweak process 
$\gamma \to \nu \bar \nu$ is suppressed by a factor $(G_F \;m^2_e)^2$ 
as compared with the probability~(\ref{eq:max}) of the pure electromagnetic 
process 
$\gamma \to e^- e^+$. However, the process with neutrinos could play a role 
of an additional channel of stellar energy-loss.

The energy-loss rate per unit volume of the photon gas due to the decay 
$\gamma \to \nu \bar \nu$ is defined by

\begin{equation}
Q =  \int \frac{d^3 k}{(2 \pi)^3} \; \frac{1}{ e^{\omega/T} - 1} \;
\omega \; W_{\perp}. 
\label{eq:Q1}
\end{equation}

\noindent Substitution of the probability~(\ref{eq:W1}) into Eq.~(\ref{eq:Q1}) 
gives

\begin{equation}
Q = \frac{\alpha \,G_F^2}{8 \pi^4}\;m_e^5 \; (e B)^2 \; 
{\cal F}(T) 
\simeq 0.96 \cdot 10^{18} \; \frac{erg}{cm^3\;s} \; 
\left (\frac{B}{B_e} \right )^2 \;{\cal F}(T).
\label{eq:Q2} 
\end{equation}

\noindent The temperature function ${\cal F}(T)$ is rather complicated in the 
general case. It takes a simple form in the limiting cases:

i) In the strong magnetic field, $e B \gg T^2 \gg m_e^2$ 

\begin{equation}
{\cal F}(T)  \simeq  \frac{T}{2 m_e}\;
\left ( \ln \frac{T}{m_e} \ln\frac{4 T}{\Gamma_\gamma} - 0.187
 \right ). 
\label{eq:F2} 
\end{equation}

\noindent Here the main contribution arises from the vicinity of the resonance 
$q^2_\parallel \simeq 4 m^2_e$. The width $\Gamma_\gamma$ should be taken from 
Eq.~(\ref{eq:max}) at $\omega = 2 m_e$.

ii) In another limit of the relatively weak magnetic field, 
$T^2 \gg e B, m_e^2$, 
one obtains 
ÿ
\begin{equation}
{\cal F}(T)  \simeq  
\frac{4 \zeta (5)}{\pi^2}\;
\left (\frac{T}{m_e} \right )^5. 
\label{eq:F3} 
\end{equation}

In a case of the low temperatures $T \ll 2 m_e$ the energy-loss rate is 
suppressed by the small factor $\exp(-2 m_e/T)$. 

In conclusion, we estimate the contribution of the photon decay 
$\gamma \to \nu \bar \nu$ into the neutrino luminosity in a supernova 
explosion, from a region of order of a hundred kilometres in size 
outside the neutrinosphere, where a rather strong magnetic field of the 
toroidal type could exist~\cite{Blinn} 

\begin{equation}
\frac{d E}{d t} \sim 10^{45} \, \frac{erg}{s} 
\left (\frac{B}{10^{15} G} \right )^2 
 \left (\frac{T}{2 MeV} \right )^5 
 \left (\frac{R}{100 km} \right )^3. 
\label{eq:lum} 
\end{equation}

\noindent It is obviously much smaller than the total neutrino luminosity 
from the neutrinosphere $ \sim 10^{52} erg/s$. It is interesting, 
however, that the process $\gamma \to \nu \bar \nu$ could give an 
appreciable contribution, equal for all neutrino species, to the low-energy 
part of the neutrino spectrum.

\bigskip

\noindent 
{\bf Acknowledgements}  

The authors are grateful to S.I.~Blinnikov  for useful discussion. 
This work was supported in part by the INTAS Grant N~96-0659.
The work of A.K. was supported in part by the ISSEP Grant N~d97-872.

\newpage
%\thispagestyle{empty}

%%%%%%%%%%%%%%%%%%%%%%%%%%%%%%  Fig.1 %%%%%%%%%%%%%%%%%%%%%%%%%%%%%%%%%%%%%

\begin{figure}[tb]

\unitlength=1.00mm
\special{em:linewidth 0.4pt}
\linethickness{0.4pt}

\begin{picture}(45.00,50.00)(-40.00,10.00)

\put(35.00,32.50){\oval(20.00,15.00)[]}
\put(35.00,32.50){\oval(16.00,11.00)[]}
\put(26.00,32.50){\circle*{2.00}}
\put(44.00,32.50){\circle*{2.00}}

\linethickness{0.8pt}

\put(59.00,42.50){\vector(-3,-2){9.00}}
\put(44.00,32.50){\line(3,2){6.00}}
\put(44.00,32.50){\vector(3,-2){9.00}}
\put(53.00,26.50){\line(3,-2){6.00}}

\put(36.50,39.00){\line(-3,2){4.01}}
\put(36.50,39.00){\line(-3,-2){4.01}}

\put(32.50,26.00){\line(3,2){4.01}}
\put(32.50,26.00){\line(3,-2){4.01}}

\put(55.00,45.00){\makebox(0,0)[cb]{\large $\nu (- p ')$}}
\put(55.00,21.00){\makebox(0,0)[ct]{\large $\nu (p)$}}
\put(34.00,45.00){\makebox(0,0)[cc]{\large $e$}}
\put(34.00,20.00){\makebox(0,0)[cc]{\large $e$}}
\put(16.00,36.00){\makebox(0,0)[cb]{\large $\gamma (q)$}}

%%%%%%%%% Definition for the photon line %%%%%%%%%%%%%%%%%%%%%%

\def\photonatomright{\begin{picture}(3,1.5)(0,0)
                                \put(0,-0.75){\tencircw \symbol{2}}
                                \put(1.5,-0.75){\tencircw \symbol{1}}
                                \put(1.5,0.75){\tencircw \symbol{3}}
                                \put(3,0.75){\tencircw \symbol{0}}
                      \end{picture}
                     }
\def\photonrighthalf{\begin{picture}(30,1.5)(0,0)
                     \multiput(0,0)(3,0){5}{\photonatomright}
                  \end{picture}
                 }

\put(10.00,32.50){\photonrighthalf}

\end{picture}

\end{figure}

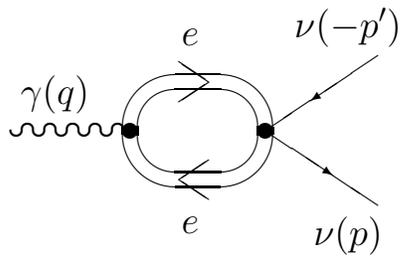
Fig.1. The Feynman diagram for the process
$\gamma \to \nu \bar \nu$. The double line corresponds 
to the exact propagator of an electron in a magnetic field.

%%%%%%%%%%%%%%%%%%%%%%%%%%%  end Fig.1 %%%%%%%%%%%%%%%%%%%%%%%%%%%%%%%%%%%%%

\end{document}